# A Thoracic Mechanism of Mild Traumatic Brain Injury Due to Blast Pressure Waves

Amy Courtney, Ph.D., Department of Physics,
United States Military Academy, West Point, NY 10996
amy_courtney@post.harvard.edu

Michael Courtney, Ph.D., Ballistics Testing Group, P.O. Box 24, West Point, NY 10996
michael_courtney@alum.mit.edu

**ABSTRACT**

The mechanisms by which blast pressure waves cause mild to moderate traumatic brain injury (mTBI) are an open question. Possibilities include acceleration of the head, direct passage of the blast wave via the cranium, and propagation of the blast wave to the brain via a thoracic mechanism. The hypothesis that the blast pressure wave reaches the brain via a thoracic mechanism is considered in light of ballistic and blast pressure wave research. Ballistic pressure waves, caused by penetrating ballistic projectiles or ballistic impacts to body armor, can only reach the brain via an internal mechanism and have been shown to cause cerebral effects. Similar effects have been documented when a blast pressure wave has been applied to the whole body or focused on the thorax in animal models. While vagotomy reduces apnea and bradycardia due to ballistic or blast pressure waves, it does not eliminate neural damage in the brain, suggesting that the pressure wave directly affects the brain cells via a thoracic mechanism. An experiment is proposed which isolates the thoracic mechanism from cranial mechanisms of mTBI due to blast wave exposure. Results have implications for evaluating risk of mTBI due to blast exposure and for developing effective protection.

**Keywords**: *traumatic brain injury, TBI, ballistic pressure wave, blast wave, blast injury, behind armor trauma, wound ballistics*

## I. INTRODUCTION

Though the concept of traumatic brain injury (TBI) due to blast wave exposure is not new (1,2), mild-to-moderate traumatic brain injury (mTBI) resulting from blast pressure waves has garnered attention as the signature injury of recent military conflicts in the Middle East (3,4).

Pressure waves can injure neural cells (5,6). However, the physical mechanisms by which blast pressure waves reach the brain and cause injury are an open question. Several hypotheses, which are not mutually exclusive, have been suggested. Possibilities include acceleration of the head, direct passage of the blast wave via the cranium, and propagation of the blast wave to the brain via a thoracic mechanism. It is important to determine how blast pressure waves are transmitted to the brain so that exposure risks can be assessed and so effective preventive measures can be implemented.

### The thoracic hypothesis

Cernak *et al.* published epidemiologic studies of mTBI due to blast injury (7,8) as well as results of several experiments in animal models investigating how blast waves may be transmitted to neural tissue and what cellular alterations result (9,10,11,12). This body of work led to a hypothesis that a thoracic mechanism may underlie mTBI due to blast exposure (13,14).

In this paper, the hypothesis that a blast pressure wave can be transmitted via a thoracic mechanism to the brain and result in cerebral effects is considered. Results of ballistic as well as blast pressure wave research are included, since ballistic and blast pressure waves, and the injuries they can cause, are similar. The ballistic pressure waves originate in the thorax, thus isolating a thoracic mechanism of propagation from cranial mechanisms. In order to link results from ballistic pressure wave research to mTBI due to blast pressure waves, an experimental design is proposed to apply a blast-like pressure wave to the thorax while blocking the cranium from direct exposure in an animal model.

*Alternate mechanisms*

The thoracic hypothesis does not exclude other mechanisms of mTBI due to blast. Sudden accelerations of the head can result in mTBI. Falls, collisions in sports, and automobile accidents are well known causes of injurious accelerations. Peak intracranial pressures due to acceleration (15,16) can be in the same range as pressure magnitudes that result in mTBI in lateral fluid percussion (LFP) models (17). In some of these studies, it is thought that damage may be caused by shear stresses at the interface of the midbrain and surrounding tissues. In addition to primary brain injury due to accelerations, blast waves can cause victims to fall or be struck by objects, resulting in secondary or tertiary injuries (18).

When the whole body is exposed to blast, it is also possible that the blast wave may reach the brain through some cranial mechanism. For example, it has been shown that blast pressure waves pass through the (thin) cranium of rats almost unchanged (19); however, it is still under investigation how a blast wave interacts with the human cranium (20).

In addition to the possibility of passing directly through the cranium, the blast wave may be entering the cranial cavity through orbital and/or aural openings. This is theoretically possible, and blast overpressure has been shown to cause direct damage to ocular neurons as deep as the midbrain in rats (21). It is unclear whether this mechanism may be contributing to mTBI since the vast majority of ocular injuries due to blast are secondary to penetration by fragments (22,23,24).

In contrast, the incidence of tympanic membrane rupture due to blast is high (9% - 47% in various studies) (22,25,26) and correlates with mTBI incidence (27,28), which suggests it may be an indicator for further evaluation. While ear protection reduces maximum pressure in the ear canal (29,30), it is unclear what role, if any, hearing protection plays in preventing mTBI due to blast.

**II. PHYSICS OF BALLISTIC PRESSURE WAVES**

The characteristics of ballistic and blast pressure waves are similar. A ballistic pressure wave is generated when a ballistic projectile enters a viscous medium (31). As the projectile loses energy over a short distance, large forces are generated that create pressure waves (force per unit area) that propagate through the medium. The magnitude of the pressure wave at a specific location is the sum of interfering waves generated at points along the projectile's path, waves generated by the collapse of the temporary cavity, and waves generated by reflections from boundaries. Pressure wave magnitude depends heavily on the projectile's local rate of energy loss.

A projectile that loses sufficient energy in a short distance in biological tissue may cause remote damage due to similar physical phenomena (32, 33, 34, 35, 36, 37, 38). The primary pressure wave (force per unit area) is generated by the projectile's loss of energy (which is the mechanical work) and propagates through the tissue. A pressure wave in the thoracic cavity will refract through and reflect from internal structures, resulting in local pressure maxima (31, 35, 37, 39). Tissue may be damaged anywhere the pressure magnitude is sufficiently large.

An anatomical example of reflected and refracted ballistic pressure waves was provided by Harvey and McMillen (40). Their high speed photographs illustrate the local maxima due to the interaction of primary and reflected waves generated by a ballistic projectile impacting a slab of beef ribs submerged in water. Sturtevant (37) presented a case study of remote damage to the spinal cord along with calculations of how the pressure wave could have reflected and refracted to produce an injurious local maximum. This phenomenon helps to explain the focal nature of internal injuries from ballistic and blast pressure waves.

Similarly, blast pressure waves reflect from the ground and nearby structures to create a complex pressure distribution that can result in localized pressures to the thorax that are several times higher than the magnitude of a blast pressure wave in a free field (41).

Other characteristics of ballistic and blast pressure waves are also similar. Peak pressures are typically reached in a few microseconds, then pressure decreases exponentially in time over a pulse duration typically less than 2 ms (31, 41). It is reasonable to expect that similar waves would cause similar damage, and similar injuries are observed, including cerebral effects (7, 11, 12, 34, 41, 42).

## III. REMOTE INJURIES DUE TO PRESSURE WAVES CAUSED BY PENETRATING BALLISTIC PROJECTILES

Injuries due to ballistic pressure waves have been reported in both case studies and experiments. A Vietnam-era database of casualties includes observations of remote wounding due to pressure waves created by penetrating injuries, including two cases of remote lung injury, five cases of remote abdominal injury in which the peritoneum was *not* perforated, and at least one example of temporary nerve damage (36). Additional case studies also show indirect injuries (43, 44, 45, 46), including indirect neural injuries (37, 47, 48), due to pressure waves created by ballistic projectiles in soft tissues.

How far from the bullet path can remote effects of ballistic pressure waves be observed? Lai *et al.* (49) used a dog model to evaluate injury to vascular endothelial cells due to distant penetration (up to approximately 0.5 m) by military rifle bullets. The number of circulating endothelial cells (a systemic indicator of damage) increased with proximity of the penetration site (leg, abdomen or thorax). Locally, laceration of vascular intima was observed in the aorta, common carotid and middle cerebral arteries by electron microscopy. These results suggest that the ballistic pressure waves traveled to the brain via the large vessels and retained sufficient magnitude to cause endothelial damage to cerebral arteries.

Suneson *et al.* (33) recorded bursts of high frequency pressure waves of average magnitude about 150 kPa (22 psi) in the brain of anesthetized pigs shot in the thigh. In this model, a mean of 728 J of work was done by the projectile. Apneic periods lasting several seconds occurred following the injury. Histological observations were limited to 'minor damage' to the blood-brain and blood-nerve barriers. Goransson *et al.* (50) reported electroencephalogram (EEG) suppression and transient apnea in pigs similarly injured. In subsequent experiments (34, 35) about 770 J of work was done by the projectile, and pressures in the range of 180-240 kPa (26-34 psi) were recorded in the brain. Microscopic damage in hippocampal and cerebellar neurons was also observed. The authors concluded that these effects were caused by pressure waves transmitted to the brain from the distant (0.5 m) point of origin.

Wang *et al.* (38) performed similar experiments in groups of dogs. In one experimental group, about 131 J of work was done by the ballistic projectile, and in the second, 740 J. Assays and electron microscopic examination showed neural damage in both groups compared with the uninjured control group; the damage in the 740 J group was more severe and detected earlier after injury. Observed damage was localized to neurons in the hippocampus. In the 740 J group, damage extended to the hypothalamus as well.

Like ballistic pressure waves, blast waves also cause focal internal injuries, including mTBI (18, 22, 28, 41, 51, 52, 53, 54). These injuries are similar to remote injuries caused by ballistic pressure waves, and may be caused by similar mechanisms. For example, in pigs exposed to blast waves, lung and intestinal injuries were observed (54). Moreover, EEG suppression, accompanied by transient apnea, was observed immediately after the blast and gradually returned to normal with 1-2 minutes.

## IV. REMOTE INJURIES DUE TO BEHIND ARMOR TRAUMA

Individual body armor is designed to prevent penetrating injuries due to ballistic projectiles. The use of body armor has thus saved many lives. At the same time, serious injuries can be sustained behind armor due to impact. As with penetrating ballistic injuries, not all impacts to body armor result in serious trauma (55, 56). However, when sufficient force reaches the tissue behind the armor, internal injuries do result (57) and may include remote damage to the central nervous system (CNS) (42, 58, 59).

For a given load and impact velocity, the deformation of body armor due to ballistic impact is a common measure of its protective ability. The mechanical work done on tissues behind armor can be estimated using information on impact energy and armor deformation (see Appendix 1). Thus conditions resulting in behind armor injuries can be compared with conditions resulting in remote injuries due to ballistic projectiles.

For example, Gryth *et al.* (59) reported severe behind armor injuries and even death in human-sized (60 kg) swine shot with 7.62 mm bullets at about 800 m/sec (from a Swedish AK4 assault rifle). Two groups, protected by armor allowing 34 mm and 40 mm of behind armor deformation, were tested. The amount of mechanical work done on the tissues was approximately 500 J and 760 J, respectively (Figure A).

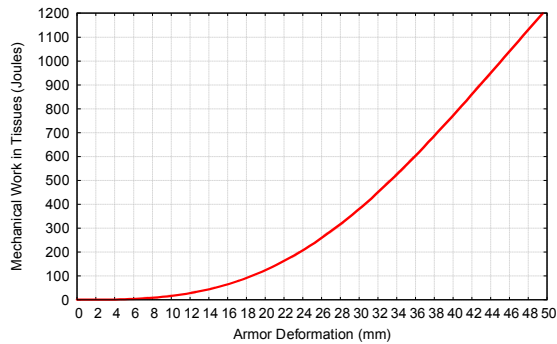

*Figure A:* Mechanical work in tissues as a function of armor deformation for behind armor trauma for impacts with 3125 J (e.g. 7.62 mm military bullet traveling 800 m/sec).

The EEG signals of two of ten test subjects in the 760 J group flatlined immediately, and five subjects died before the end of the two-hour observation period. Two of eight test subjects in the 500 J group also died within two hours. Surviving subjects in both groups experienced EEG suppression that usually resolved within a few minutes. All test subjects showed behind armor trauma consisting of lung hematoma; additional internal injuries included rib fracture, emphysema and hemoptysis.

In another test of seven large swine protected by armor allowing 28 mm of deformation (320 J of mechanical work done by the projectile on thoracic tissue), all subjects survived but suffered unilateral lung hematomas (60). In five of seven test subjects a reduction in EEG activity occurred after the shot. In each case, EEG activity returned to baseline within two minutes. These EEG changes are similar to those observed by Goransson *et al.* (50) following penetrating injuries of the hind limbs of pigs and by Axelsson *et al.* (54) following blast wave exposure in pigs.

Roberts *et al.* performed physical and finite element modeling of behind armor trauma in humans (61, 62, 63). They concluded that while armor meeting National Institutes of Justice standards may protect from penetrating injuries, it may not protect from behind armor trauma. Their calculations predict that pressure waves initiated when sufficient energy is transmitted through body armor are similar to those produced by penetrating ballistic projectiles and blast waves (63). Therefore, it is not surprising that individual body armor is not always effective against behind armor trauma or blast wave injury (64, 65). Work is ongoing to develop individual body armor that absorbs more energy, thus allowing less to be transmitted to the thorax (66, 67).

## V. DISCUSSION

### *How much pressure is damaging to neurons?*

It is important to quantify how much pressure is damaging to neurons, and under what conditions ballistic or blast waves might reach the brain with damaging results. However, direct pressure measurements are not available for every experiment, and researchers have characterized ballistic or blast conditions differently. For ballistic pressure waves (penetrating or behind armor), the local rate of mechanical work governs the magnitude of the pressure wave in tissue. Therefore, characterizing ballistic pressure waves in terms of the mechanical work done by the projectile allows comparison between results of penetrating and behind-armor experiments.

Local pressure wave magnitudes associated with mild, moderate, and severe TBI have been determined using the lateral fluid percussion (LFP) model of brain injury (17). Mild and moderate brain injuries occurred with local pressure levels in the range of 100-300 kPa (15-45 psi). Pressure waves near 200 kPa (30 psi) caused immediate incapacitation in laboratory animals in one study (68).

Pressure wave magnitudes in this range were measured in animal models of penetrating ballistic injuries in which approximately 700 - 800 J of work was done on tissues 0.5 m distant from the brain (33, 34, 35). Localization of damage to neurons in the hippocampus and hypothalamus was also observed. These results were supported by similar, independent experiments in dogs (38). The degree of remote neural damage was dependent on the mechanical work done by the projectile, and was even detectable at about 130 J, which might be expected to produce pressure wave magnitudes of only 55 kPa (7-8 psi) in the brain.

Localization of observed neural damage to the hippocampus and the hypothalamus agrees with the results of experiments using the lateral fluid percussion (LFP) model of traumatic brain injury

(17) as well as clinical studies (69, 70, 71, 72). The results of the independent experiments discussed in section III agree that a pressure wave generated by a penetrating projectile at a distant location can reach the brain with enough magnitude to cause neuronal damage in the same region of the brain most affected in mTBI.

In summary, when the mechanical work done by the projectile is 700-800 J, it is probable that transient pressure wave magnitudes of 15-45 psi occur in the brain. The likelihood that a bullet impact remote from the brain will result in mTBI in humans can also be expected to increase with the mechanical work applied to the tissue (73). It has been estimated that when this much work is applied to the thorax, half of human victims will likely experience rapid incapacitation and show signs of mTBI.

### *Are observed cerebral effects vagally mediated?*

Some have suggested that CNS effects due to blast or ballistic pressure waves might not be due to direct exposure of brain tissue to the pressure wave, but might instead be mediated by the vagus nerve (9, 42, 74). Experimental results indicate that a vagally-mediated response is present, but that if the pressure wave reaches the brain with sufficient magnitude, it will cause damage. In human-sized pigs, periods of apnea were observed in intact animals when about 700 J of work was done on tissue behind body armor. Test subjects on which bilateral vagotomy had been performed did not exhibit apnea. However, vagotomy mitigated but *did not eliminate* EEG suppression one minute after the ballistic impact. Both experimental groups showed evidence of lung contusions due to behind armor trauma.

A limitation of this study is that EEG records are reported at one-minute intervals, while important information may be contained in data from the first minute. In the ballistic pressure wave experiments by Goransson *et al.* (50) as well as the blast pressure wave experiments by Axelsson *et al.* (54), EEG changes that may be important indicators of neural effects were observed to occur "immediately" after the blast and were no longer present after one or two minutes.

Irwin *et al.* (74) also observed that rats exposed to blast experienced bradycardia and hypotension - effects that were mitigated in a group on which bilateral vagotomy had been performed. Respiratory and EEG data were not reported.

Both experimental groups sustained severe blast lung injury. Cernak *et al.* reported similar results in rabbits exposed to a focused blast wave applied to the middle thoracic region. Vagotomy reduced but *did not eliminate* damage directly to brain cells (9).

The results of these animal studies of ballistic and blast wave exposure suggest that the vagus nerve does play a role in the CNS response to pressure waves. In particular, apnea and bradycardia after exposure seem to be vagally-mediated responses. The resulting hypoxia may exacerbate the consequences of neural cell damage. However, the results of these experiments also indicate that brain cells can be damaged directly after exposure to pressure waves originating or focused at a distant location.

A potential limitation of Cernak *et al.*'s blast wave experiment on rabbits is the inability to truly restrict the blast wave to the thoracic region (due to refraction of the blast wave at the edge of the shock tube). One is not certain of the magnitude of the external pressure wave applied to the rabbit's head in these experiments.

### *Underwater experiments*

Knudsen and Oen (75) were in a unique position to observe effects of blast waves that originate internally. In Norway, minke whales are (legally) hunted from small fishing boats using grenade-tipped harpoons. The harpoon is aimed at the thorax, and the grenade detonates 60-70 cm inside the animal, which is about 10 m long and 8500 kg at maturity. In a sample of 37 whales, the degree of brain injury was greater for detonations occurring closer to the brain, as one might expect. Interestingly, fatal neurotrauma was also caused by detonations as far back as the rostral abdomen and in the absence of gross trauma to the brain (microscopic trauma was evident). This experiment involved an extreme situation in which neurotrauma resulted in death rather than mTBI. However, it clearly demonstrates that blast waves can be transmitted to the brain via the thorax and result in neural damage.

Military case reports indicate that primary blast injury and mortality are greater when the blast and victims are under water (64). This is due to the greater coupling of the blast wave to the body, which has an acoustic impedance close to that of water[1]. Phillips and Richmond (64) related qualitative results of a relevant German study on dogs after World War II. Dogs submerged except for the head and exposed to underwater blast experienced internal injuries "pathologically identical to that of air blast." When the head alone was submerged, no internal organ injuries were observed, suggesting that the water/air interface was an effective barrier. Unfortunately, neurological effects are not discussed. This result is supported by those of Yelverton *et al.* (77), who reported internal injuries in submerged waterfowl exposed to under water blast but not in waterfowl floating on the water's surface.

### *A proposed experiment*

An experiment in which an externally generated, nonlethal pressure wave could be isolated from the cranium would demonstrate whether a blast pressure wave can be transmitted to the brain via the thorax. The above results suggest that a design in which the test subject is submerged except for the head while a blast-like wave is generated under water would isolate the cranium from the direct pressure wave.

In an unpublished experiment (78), a blast-like pressure wave was generated by firing a ballistic projectile vertically into water near a terrestrial animal model. The path of the projectile was very near (6-8 cm) but did not penetrate the ventral thorax of each test subject, which was submerged except for the head. Immediate incapacitation and death were observed in several test subjects in the absence of any penetrating wounds. Incidence of incapacitation and death increased with the magnitude of the blast-like pressure wave.

A similar experimental model can be used to assess mTBI due to blast pressure waves applied to the thorax (Figure B). Protocols for physiological monitoring and mTBI detection are established, so only a few comments are included here.

Key measurements include the pressure at four locations: immediately outside the subject's thorax, inside the thorax, inside the brain, and outside the head (in air, to verify that the head is not directly exposed to the pressure wave). High speed pressure sensors (19) are needed for accurate signal acquisition.

If test subjects are anesthetized, it is important to note that immediate behavioral observations will be precluded, and EEG measurements will be affected (79). EEG signals are known to decrease with depth of anesthesia, so quantitative analysis of EEG data would need to take that into consideration. However, anesthesia does not seem to affect cardiovascular responses or neural damage.

It is essential that the blast-like pressure wave be initiated under water. However, several means are likely acceptable, such as firing a ballistic projectile into water, detonating an explosive charge under water, or perhaps using an under water shock tube. We favor a ballistic method because the energy at impact and distance from the test subject are easily controlled. A firing distance of about 3 m above the water surface will prevent the muzzle blast from interfering with the desired pressure wave. A metal plate placed near the firing mechanism and with a small hole for passage of the projectile (32, 33, 34, 35) will further shield the cranium from possible exposure to the muzzle blast.

A simple estimate of the peak pressure (*p*) a distance (*r*) from the bullet path is

$$p \approx \frac{10E}{d\,4\pi r^2} \quad (1),$$

where *E* is the kinetic energy of the bullet (one half of the mass times the square of the speed), and *d* is the penetration distance (73). Note that the magnitude of the pressure wave decreases as the square of the distance from the bullet path so accurate distance measurements are necessary (Figure C).

This formula is approximate. Also, while the acoustic impedances of water and soft tissue are similar, the exact transfer function for the pressure wave is not known and will vary some with the animal model used. Therefore, preliminary tests must be performed to verify actual pressures in the specific experimental design.

---

[1] The acoustic impedance of air is approximately $0.40\,\mathrm{kPa\cdot s/m}$, while the acoustic impedance of water is approximately $1500\,\mathrm{kPa\cdot s/m}$ and of human tissues averages $1500\text{-}1700\,\mathrm{kPa\cdot s/m}$ (76), respectively. Physically this means that a blast wave initiated in air can penetrate water, but that almost none of a blast wave initiated under water is transmitted to air.

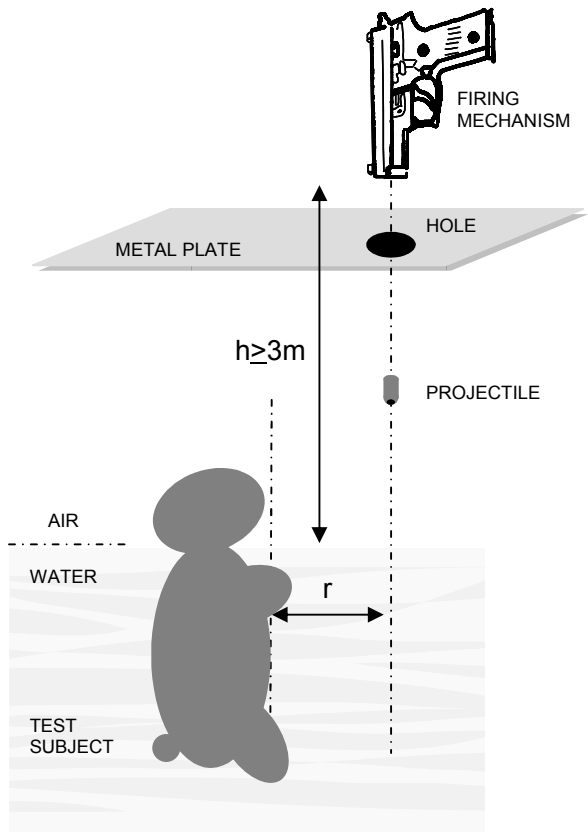

*Figure B.* A possible arrangement of animal test subject and projectile path, not shown to scale. The distance r between the projectile path and the surface of the thorax is used to estimate the magnitude of the pressure wave incident on the thorax. The distance h and the metal plate shield the cranium from possible muzzle blast exposure. High speed pressure and physiologic sensors are not shown.

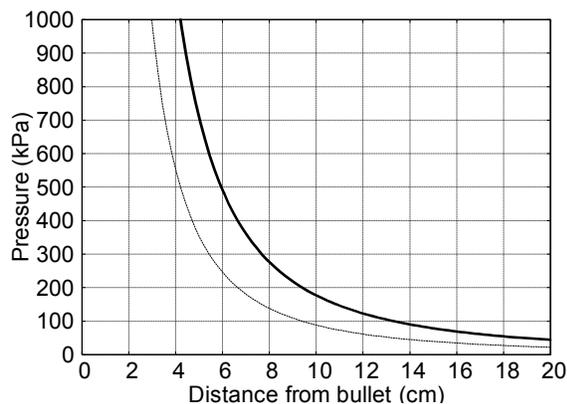

*Figure C*: Estimates of peak pressure magnitude reaching a test subject for the experimental set-up shown in Figure B. The solid line is for an expanding (hollow point) bullet shot from a .40 S&W pistol. The dashed line is for a non-expanding (full metal jacket) bullet.

The water/air boundary will shield the cranium from external exposure to the pressure wave. Though brain injury is possible at lower transient pressure levels, we predict that a pressure wave of magnitude 200-400 kPa (about 30-60 psi) inside the thorax will produce observable cerebral effects.

## VI. CONCLUSION

Possible mechanisms by which blast pressure waves cause mild traumatic brain injury (mTBI) include acceleration of the head, direct passage of the blast wave via a cranial mechanism, and propagation of the blast wave to the brain via a thoracic mechanism. In light of ballistic as well as blast pressure wave research, we considered the hypothesis that blast pressure waves may be transmitted to the brain via the thorax with enough magnitude to cause mTBI.

Studies have shown that ballistic pressure waves, resulting from penetrating ballistic projectiles or from ballistic impacts to body armor, can be transmitted to the brain from the thorax and result in injury to neural tissue. Characterizing experimental conditions in terms of the mechanical work done by the ballistic projectile permits comparisons across studies. This analysis suggests that when the mechanical work done on thoracic tissue is around 700 J, neural injury will likely result. Neural damage has been detected at lower levels of mechanical work in an animal model. However, the lower threshold for clinically significant injury is unclear.

Blast waves applied to the whole body or focused on the thorax of animal models have been shown to cause cerebral effects similar to those caused by ballistic pressure waves. These effects include EEG suppression and damage to hippocampal and hypothalamic neurons, areas of the brain that have been associated with mTBI in humans.

Taken together, these results provide strong support for a thoracic mechanism of mTBI due to blast wave exposure. However, in the blast wave experiments in animal models, even when the blast wave was focused on the thorax, it is not known how much of the blast wave reached the brain via a thoracic mechanism rather than some other mechanism.

Therefore, an experiment has been proposed in which a blast-like pressure wave is generated under water and transmitted to the thorax but blocked from the cranium in an animal model. If no neural damage is observed when the blast-like wave is of a magnitude expected to cause neural injury, the thoracic hypothesis would not be supported.

## APPENDIX 1

### Behind armor retarding force and mechanical work in tissues from projectile impact

Several papers (42, 59, 60) reported remote cerebral effects due to projectiles stopped by armor over the thorax. Severity of both local injuries and remote cerebral effects increased with the deformation of the body armor during impact, which is measured as the impression left when the body armor is impacted with a plasticine backing. Behind armor trauma is often parameterized in terms of the impact energy and armor deformation. Here, simple methods of approximating the behind armor mechanical work and resulting retarding force are developed to facilitate comparison with remote effects of penetrating projectiles and blast injuries.

A simple model of behind armor mechanical work, *W*, as a function of armor deformation, *d*, is

$$W(d) = E \times \left( 1 - \frac{1}{1 + \left(\frac{d}{d_0}\right)^3} \right),$$

where *E* is the impact energy of the bullet, and $d_0$ is an adjustable parameter that sets the armor deformation depth at which half of the impact energy is available to do work creating forces and pressures in behind armor tissues.

This model is physically reasonable. It has the expected monotonic and limiting behaviors, giving zero work for body armor which stops the projectile with zero armor deformation and approaching the maximum possible work as the armor deformation becomes very large. Work done by penetrating projectiles is approximately proportional to the volume of displaced tissue (temporary cavitation) (80). For small deformations, the volume of displaced tissue will scale as the deformation cubed, so the model gives the expected scaling at small displacements.

Several papers report behind armor trauma for projectile impact energy of approximately 3125 J and armor deformations ranging from 28 to 44 mm (42, 59, 60). Comparing their descriptions of resulting lung injuries with experiments of penetrating handgun bullets in deer (81, and Courtney and Courtney, unpublished data) suggests a value of $d_0$ = 58 mm. This value of $d_0$ gives agreement between the level of lung damage and the work done in the tissues between the behind armor trauma experiments and the deer experiments where the value of mechanical work in tissues is independently known. The resulting mechanical work done by the projectile on tissues for a specific case is shown in Figure A.

The average force that the bullet exerts on tissues behind armor can then be computed, because the average force is simply the mechanical work divided by the stopping distance, *d*. The average retarding force for the condition considered above is shown in Figure D. Note that the average retarding force in tissue for armor deformation of 40 mm is close to 19,000 N, which is comparable to the peak retarding force of a penetrating 7.62 mm bullet at full yaw (36).

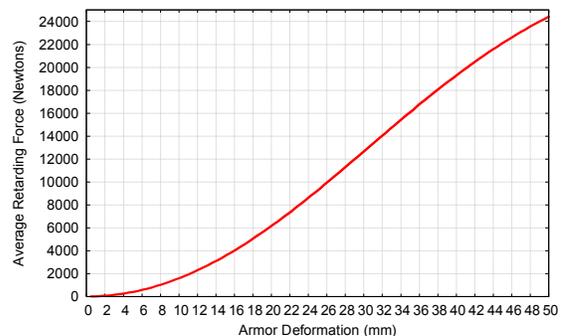

*Figure D:* Average retarding force in tissues as a function of armor deformation for behind armor trauma for impacts with 3125 J.

**References**

1. Sharp D. Shocked, shot, and pardoned. Lancet 2006;368:975-976.

2. Jones E, Fear NT, Wessely S. Shell shock and mild traumatic brain injury: a historical review. Am J Psych 2007;164:1641-1645.

3. Hoge CW, McGurk D, Thomas JL, Cox AL, Engel CC, Castro CA. Mild traumatic brain injury in U.S. soldiers returning from Iraq. New Engl J Med 2008;358(5):453-463.

4. Moore DF, Radovitzky RA, Shupenko L, Klinoff A, Jaffee MS, Rosen JM. Blast physics and central nervous system injury. Future Neurol 2008; 3(3):243-250.

5. Saljo A, Huang Y, Hansson H. Impulse noise transiently increased the permeability of nerve and glial cell membranes, an effect accentuated by a recent brain injury. J Neurotrauma 2003;20(8): 787-794.

6. Suneson A, Hansson HA, Lycke E, Seeman T. Pressure wave injuries to rat dorsal root ganglion cells in culture caused by high energy missiles. J Trauma 1989;29(1):10-18.

7. Cernak I, Savic J, Ignjatovic D, Jevtic M. Blast injury from explosive munitions. J Trauma 1999;47(1):96-103.

8. Cernak I, Savic J, Zunic G, Pejnovic N, Jovanikic O, Vladislav S. Recognizing, scoring and predicting blast injuries. World J Surg 1999;23:44-53.

9. Cernak I, Savic J, Malicevic Z, Zunic G, Radosevic P, Ivanovic I, Davidovic L. Involvement of the central nervous system in the general response to pulmonary blast injury. J Trauma 1996;40(3S):S100-S104.

10. Cernak I, Savic J, Malicevic Z, Zunic G, Radosevic P, Ivanovic I. Leukotrienes in the pathogenesis of pulmonary blast injury. J Trauma 1996;40(3S):S148-S151.

11. Cernak I, Wang Z, Jiang J, Bian X, Savic J. Cognitive deficits following blast-induced neurotrauma: possible involvement of nitric oxide. Brain Inj 2001;15(7):593-612.

12. Cernak I, Wang Z, Jiang J, Bian X, Savic J. Ultrastructural and functional characteristics of blast injury-induced neurotrauma. J Trauma 2001;50(4):695-706.

13. Cernak I. Penetrating and blast injury. Rest Neuro Neurosci 2005;23:139-143.

14. Bhattacharjee Y. Shell shock revisited: solving the puzzle of blast trauma. Science 2008;319:406-408.

15. Zhang L, Yang K, King A. A proposed injury threshold for mild traumatic brain injury. J Biomech Eng 2004;126:226-236.

16. Krave U, Hojer S, Hansson H. Transient, powerful pressures are generated in the brain by a rotational acceleration impulse to the head. Euro J Neurosci 2005;21:2876-2882.

17. Thompson HJ, Lifshitz J, Marklund N, *et al*. Lateral fluid percussion model of brain injury: a 15-year review and evaluation. J Neurotrauma 2005;22(1):42-75.

18. Finkel MF. The neurological consequences of explosives. J Neurol Sci 2006;249:63-37.

19. Chavko M, Koller WA, Prusaczyk WK, McCarron RM. Measurement of blast wave by a miniature fiber optic pressure transducer in the rat brain. J Neurol Meth 2007;159:277-281.

20. Singer E. Brain trauma in Iraq. Technology Review 2008;111(3):52-59.

21. Petras JM, Bauman RA, Elsayed NM. Visual system degeneration induced by blast overpressure. Toxicology 1997;121(1):41-49.

22. Karmy-Jones R, Kissinger D, Golocovsky M, Jordan M, Champion HR. Bomb-related injuries. Mil Med 1994;159:536-539.

23. Thach AB, Johnson AJ, Carroll RB, et al. Severe eye injuries in the war in Iraq, 2003-2005. Ophthalmology 2008;15(2):377-382

24. Mines M, Thach A, Mallonee S, Hildebrand L, Shariat S. Ocular injuries sustained by survivors of the Oklahoma City bombing. Ophthalmology 2000;107(5):837-843.

25. Sylvia FR, Drake AI, Wester DC. Transient vestibular balance dysfunction after primary blast injury. Mil Med 2001;166(10):918-920.

26. Gutierrez de Ceballos JP, Turegano-Fuentes F, Diaz DP, Sanchez MS, Llorente CM, Guerrero-Sanz JE. Casualties treated at the closest hospital in the Madrid, March 11, terrorist bombings. Crit Care Med 2005;33(1):S107-S112.

27. Xydakis MS, Bebarta VS, Harrison CD, Conner JC. Tympanic-membrane perforation as a marker of concussive brain injury in Iraq. New Engl J Med 2007;357(8):830.

28. Ritenour AE, Wickley A, Ritenour JS, Driete BR, Blackbourne LH, Holcomb JB, Wade CE. Tympanic membrane perforation and hearing loss from blast overpressure in Operation Enduring Freedom and Operation Iraqi Freedom wounded. J Trauma 2008;64(S2):174S-178S.

29. Ylikoski M,k Pekkarinen JO, Starck JP, Paakkonen RJ, Ylikoski JS. Physical characteristics of gunfire impulse noise and its attenuation by hearing protectors. Scand Audiol 1995;24(1):3-11.

30. Li C, Zhu P, Liu Z, Wang Z, Yang C, Chen H, Ning X, Zhou J. Comparative observation of protective effects of earplug and barrel on auditory organs of guinea pigs exposed to experimental blast underpressure. Chinese J Traumatol 2006; 9(4):242-245.

31. Lee M, Longoria RG, Wilson DE. Ballistic waves in high-speed water entry. J Fluids Structures 1997;11:819-844.

32. Suneson A, Hansson HA, Seeman T. Peripheral high-energy missile hits cause pressure changes and damage to the nervous system: experimental studies on pigs. J Trauma 1987;27(7):782-789.

33. Suneson A, Hansson HA, Seeman T. Central and peripheral nervous damage following high-energy missile wounds in the thigh. J Trauma 1988;40(S3):197S-203S.

34. Suneson A, Hansson HA, Seeman T. Pressure wave injuries to the nervous system caused by high-energy missile extremity impact: part I. local and distant effects on the peripheral nervous system. A light and electron microscopic study on pigs. J Trauma 1990;30(3):281-294.

35. Suneson A, Hansson HA, Seeman T. Pressure wave injuries to the nervous system caused by high energy missile extremity impact: part II. distant effects on the central nervous system. A light and electron microscopic study on pigs. J Trauma 1990;30(3):295-306.

36. Bellamy RF, Zajtchuk R, eds. Conventional Warfare: Ballistic, Blast and Burn Injuries. Textbook of Military Medicine 1990; Office of the Surgeon General, Department of the Army, United States of America.

37. Sturtevant B. Shock wave effects in biomechanics. Sadhana 1998;23:579-596.

38. Wang Q, Wang Z, Zhu P, Jiang J. Alterations of the myelin basic protein and ultrastructure in the limbic system and the early stage of trauma-related stress disorder in dogs. J Trauma 2004;56(3):604-610.

39. Pompei A, Sumbatyan MA, Boyev NV. Reflection of high-frequency elastic waves from a non-plane boundary surface of the elastic medium. J Sound Vibration 2007;302:925-935.

40. Harvey EN, McMillen JH. An experimental study of shock waves resulting from the impact of high velocity missiles on animal tissues. J Exper Med 1947;85:321-341.

41. Stuhmiller JH. Blast injury: translating research into operational medicine. In: Santee WR, Friedl KE, eds. Quantitative Physiology: Problems and Concepts in Military Operational Medicine. Textbook of Military Medicine 2008. Office of the Surgeon General, Department of the Army, United States of America.

42. Gryth D, Rocksen D, Arborelius UP, *et al*. Bilateral vagotomy inhibits apnea and attenuates other physiological responses after blunt chest trauma. J Trauma 2008;64(6):1420-1426.

43. Johansson L, Holmstrom A, Lennquist S, Norrby K, Nystrom PO. Intramural haemorrhage of the intestine as an indirect effect of missile trauma. Acta Chir Scand 1982;148(1):15-19.

44. Sasaki LS, Mittal UK. Small bowel laceration from a penetrating extraperitoneal gunshot wound: a case report. J Trauma 1995; 39:602-604.

45. Velitchkov NG, Losanoff JE, Kjossev KT, Katrov ET, Mironov MB, Losanoff HE. Delayed small bowel injury as a result of penetrating extraperitoneal high-velocity ballistic trauma to the abdomen. J Trauma 2000;48(1):169-170.

46. Tien HC, van der Hurk TWG, Dunlop P, *et al*. Small bowel injury from a tangential gunshot wound without peritoneal penetration: a case report. J Trauma 2007;62:762-764.

47. Treib J, Haass A, Grauer MT. High-velocity bullet causing indirect trauma to the brain and symptomatic epilepsy. Mil Med 1996;161(1):61-64.

48. Bhatoe HS, Singh P. Missile injuries of the spine. Neurol India 2003;51(4):507-511.

49. Lai X, Liu Y, Wang J, Li S, chen L, Guan Z. Injury to vascular endothelial cells and the change of plasma endothelin level in dogs with gunshot wounds. J Trauma 1996;40(3S):S60-S62.

50. Goransson AM, Ingvar DH, Kutyna F. Remote cerebral effects on EEG in high-energy missile trauma. J Trauma 1988;28(Suppl):204-205.

51. DePalma RG, Burris DG, Champion HR, Hodgson MJ. Blast injuries. New Engl J Med 2005;352:1335-1342.

52. Tatic V, Ignjatovic D, Jevtic M, Jovanovic M, Draskovic M, Durdevic D. Morphologic characteristics of primary nonperforative intestinal blast injuries in rats and their evolution to secondary perforations. J Trauma 1996;40(3S):S94-S99.

53. Yang Z, Wang Z, Tang C, Ying Y. Biological effects of weak blast waves and safety limits for internal organ injury in the human body. J Trauma 1996;40(S3):81S-84S.

54. Axelsson H, Hjelmqvist H, Medin A, Persson J, Suneson A. Physiological changes in pigs exposed to a blast wave from a detonating high-explosive charge. Mil Med 2000;165(2):119-126.

55. Carroll AW, Soderstrom CA. A new nonpenetrating ballistic injury. Ann Surg 1978;188(6):753-757.

56. Metker LW, Prather RN, Coon PA, Swann CL, Hopkins CE, Sacco WJ. A method of soft body armor evaluation: cardiac testing. Technical Report ARCSL TR78034, Aberdeen Proving Ground, Maryland, 1978.

57. Liden E, Berlin R, Janzon B, Schantz B, Seeman T. Some observations relating to behind-body armour blunt trauma effects caused by ballistic impact. J Trauma 1988;27(1S):S145-S148.

58. Cannon L. Behind armour blunt trauma – an emerging problem. J R Army Med Corps 2001;147(1):87-96.

59. Gryth D, Rocksen D, Persson JKE, *et al*. Severe lung contusion and death after high-velocity behind-armor blunt trauma: relation to protection level. Mil Med 2007;172(10):1110-1116.

60. Drobin D, Gryth D, Persson JKE, *et al*. Electroencephalogram, circulation and lung function after high-velocity behind armor blunt trauma. J Trauma 2007; 63(2):405-413.

61. Roberts JC, O'Connor JV, Ward EE. Modeling the effect of nonpenetrating ballistic impact as a means of detecting behind-armor blunt trauma. J Trauma 2005;58(6):1241-1251.

62. Roberts JC, Ward EE, Merkle AC, O'Connor JV. Assessing behind armor blunt trauma in accordance with the National Institute of Justice standard for personal body armor protection using finite element modeling. J Trauma 2007;62(5):1127-1133.

63. Merkle AC, Ward EE, O'Connor JV, Roberts JC. Assessing behind armor blunt trauma (BABT) under NIJ Standard 0101.04 conditions using human torso models. J Trauma 2008;64(6):1555-1561.

64. Phillips YY, Richmond DR. Primary blast injury and basic research: a brief history. In Bellamy RF, Zajtchuk R, eds. Conventional Warfare: Ballistic, Blast and Burn Injuries. Textbook of Military Medicine 1990. Office of the Surgeon General, Department of the Army, United States of America.

65. Cooper GJ. Protection of the lung from blast overpressure by thoracic stress wave decouplers. J Trauma 1996;40(3S):105S-110S.

66. Daraio C, Nesterenko VF, Herbold EB, Jin S. Energy trapping and shock disintegration in a composite granular medium. Phys Rev Lett 2006;96(5):580-582.

67. Doney R, Sen S. Decorated, tapered and highly nonlinear granular chain. Phys Rev Lett 2006;97(15):1555-1557.

68. Toth Z, Hollrigel G, Gorcs T, and Soltesz I. Instantaneous perturbation of dentate interneuronal networks by a pressure wave transient delivered to the neocortex. J Neurosci 1997;17(7);8106-8117.

69. Lowenstein DH, Thomas MJ, Smith DH, McIntosh TK:. Selective vulnerability of dentate hilar neurons following traumatic brain injury: a potential mechanistic link between head trauma and disorders of the hippocampus. J Neurosci 1992;12:4846-4853.

70. Herrmann M, Curio N, Jost S, *et al*. Release of biochemical markers of damage to neuronal and glial brain tissue is associated with short and long term neuropsychological outcome after traumatic brain injury. J Neurol Neurosurg Psych 2001;70:95-100.

71. Umile EM, Sandel ME, Alavi A, Terry CM, Plotkin RC. Dynamic imaging in mild traumatic brain injury: support for the theory of medial temporal vulnerability. Arch Phys Med Rehab 2002;83(11):1506-1513.

72. Metting Z, Rodiger L, Keyser J, van der Naalt J. Structural and functional neuroimaging in mild-to-moderate head injury. Lancet Neurol 2007;6(8):699-710.

73. Courtney A, Courtney M. Links between traumatic brain injury and ballistic pressure waves originating in the thoracic cavity and extremities. Brain Inj 2007;21(7):657-662.

74. Irwin RJ, Lerner MR, Bealer JF, Mantor PC, Brackett DJ, Tuggle DW. Shock after blast wave injury is caused by a vagally mediated reflex. J Trauma 1999;47(1):105-110.

75. Knudsen SK, Oen EO. Blast-induced neurotrauma in whales. Neurosci Res 2003;46:377-386.

76. Ludwig GD. The velocity of sound through tissues and the acoustic impedance of tissues. J Acoustical Soc Am 1950; 22(6):862-866.

77. Yelverton JT, Richmond DR, Fletcher ER, Jones RK. Safe distances from underwater explosions for mammals and birds. Defense Nuclear Agency Report. 1973; DNA 3114T. Accession number AD766952.

78. Courtney M, Courtney A. Experimental observations of incapacitation via ballistic pressure wave without a wound channel. http://www.ballisticstestinggroup.org/lotor.pdf accessed 24 July, 2008.

79. Shaw NA. The neurophysiology of concussion. Prog Neurobiol 2002;67:281-344.

80. Peters CE. A mathematical-physical model of wound ballistics. J Trauma (China) 1990;6(2) Supplement:308-318.

81. Courtney M, Courtney A. A method for testing handgun bullets in deer. 2007 http://arxiv.org/ftp/physics/papers/0702/0702107.pdf Accessed 24 July, 2008.